\documentclass[aps,floats,superscriptaddress,twocolumn,pre,showpacs]{revtex4-1}

\usepackage{graphicx,epsfig}
\usepackage{graphics,dcolumn,bm,epic, eepic,fleqn,float}
\usepackage{amssymb,amsmath,amsfonts,multirow,rotate,color}
\usepackage{soul}
\usepackage{wasysym}
\usepackage{dcolumn}
\usepackage{bm}

\definecolor{Myorange}{cmyk}{0,0.42,1,0}

\newcommand{\avg}[1]{\langle #1 \rangle}

\newcommand{\lay}[1]{^{[#1]}}

\begin{document}

\title{Emergence of Multiplex Communities in Collaboration Networks}

\author {Federico Battiston*}
\affiliation{School of Mathematical Sciences, Queen Mary University of London, 
London E1 4NS, United Kingdom}   

\author {Jacopo Iacovacci*}
\affiliation{School of Mathematical Sciences, Queen Mary University of London, 
London E1 4NS, United Kingdom}   

\author {Vincenzo Nicosia}
\affiliation{School of Mathematical Sciences, Queen Mary University of London, 
London E1 4NS, United Kingdom}  

\author {Ginestra Bianconi}
\affiliation{School of Mathematical Sciences, Queen Mary University of London, 
London E1 4NS, United Kingdom}  

\author{Vito Latora}
\affiliation{School of Mathematical Sciences, Queen Mary University of London, 
London E1 4NS, United Kingdom}  
\affiliation{Laboratorio sui Sistemi Complessi, Scuola Superiore di Catania, 
I-95123 Catania, Italy}
\affiliation{Dipartimento di Fisica ed Astronomia, Universit\`a di Catania and INFN, I-95123 Catania, Italy}

\date{\today}
\begin{center}
\begin{abstract}
Community structures in collaboration networks reflect the natural
tendency of individuals to organize their work in groups in order to
better achieve common goals. In most of the cases, individuals exploit
their connections to introduce themselves to new areas of interests,
giving rise to multifaceted collaborations which span different
fields. In this paper, we analyse collaborations in science and among
movie actors as multiplex networks, where the layers represent
respectively research topics and movie genres, and we show that
communities indeed coexist and overlap at the different layers of such
systems. We then propose a model to grow multiplex networks based on
two mechanisms of intra and inter-layer triadic closure which mimic
the real processes by which collaborations evolve. We show that our
model is able to explain the multiplex community structure observed
empirically, and we infer the strength of the two underlying social
mechanisms from real-world systems. Being also able to correctly
reproduce the values of intra-layer and inter-layer assortativity
correlations, the model contributes to a better understanding of the
principles driving the evolution of social networks.
\end{abstract}
\end{center}

\maketitle


\section*{Introduction}
More often than not the agents of a social system prefer to combine
their efforts in order to achieve results that would be otherwise
unattainable by a single agent alone. A relevant role in the
organisation of such systems is therefore played by the emerging
patterns of collaboration within a group of individuals, which have
been widely and thoroughly investigated in the last few
decades~\cite{wasserman1994social,john2000social}. In a collaboration network, two
individuals are considered to be linked if they are bound by some form
of partnership.  For instance, in the case of scientific
collaborations, the nodes of the networks correspond to scientists and
the relationship between two authors is testified by the fact that
they have co-authored one or more papers~\cite{newman2001structure}. Another
well-known example of collaboration network is that of co-starring
graphs, where the nodes represent actors and there is a link between
two actors if they have appeared in the same movie.

The study of large collaboration systems has revealed the presence of
a surprisingly high number of triangles in the corresponding
networks~\cite{watts1998collective,ramasco2004self}. This indicates that two nodes with a
common neighbour have a higher probability to be linked than two
randomly chosen nodes. This effect, known as
\textit{transitivity}~\cite{wasserman1994social}, can be easily explained in
terms of a basic mechanism commonly referred to as {\em triadic
  closure}~\cite{rapoport1953spread}, according to which two individuals of a
collaboration network have a high probability to connect after having
been introduced to each other by a mutual
acquaintance~\cite{newman2001clustering,watts1998collective,lee2010complete}. Some other works have
pointed out that triadic closure can also explain other empirical
features of real-world collaboration networks, including fat-tailed
degree distributions and correlations between the degrees of
neighbouring nodes~\cite{holme2002growing,bianconi2014triadic}.

Another remarkable feature often observed in social and collaboration
networks is the presence of meso-scale structures in the form of {\em
  communities}, i.e. groups of tightly connected nodes which are
loosely linked to each other~\cite{girvan2002community}. Interestingly, structural
communities quite often correspond to functional
groups~\cite{fortunato2010community}.

An important observation is that not all the links of a collaboration
network are equal, since collaborations can often be classified into a
number of different categories. For instance, scientific co-authorship
can be classified according to the research field, while actors often
appear in movies of different genres. In these cases, a collaboration
network is better described in terms of a {\em multi-layer} or {\em
  multiplex network}~\cite{kivela2014multilayer,boccaletti2014structure} where links
representing collaborations of a specific kind are embedded on a
separate layer of the network, and each layer can have in general a
different topology. Great attention has been recently devoted to the
characterisation of the structure~\cite{cardillo2013emergence,bianconi2013statistical,battiston2014structural,de2015ranking,nicosia2015measuring,de2015structural} and dynamics~\cite{gomez2013diffusion,cozzo2013contact,radicchi2013abrupt, battiston2015biased} of multi-layer networks. In particular,
various models to grow multiplex networks have appeared in the
literature, focusing on linear~\cite{nicosia2013growing} or
non-linear~\cite{nicosia2014nonlinear} preferential attachment, or on weighted
networks~\cite{murase2014multilayer}. Less attention has been devoted to define
and extract communities in multiplex
networks~\cite{de2015identifying,iacovacci2015mesoscopic}, for instance by mean of
stochastic block models~\cite{valles2014multilayer,peixoto2015inferring}.

In this work we investigate the multiplex nature of communities in
collaboration networks and we propose a simple model to explain the
appearance, coexistence and co-evolution of communities at the
different layers of a multiplex. Our hypothesis is that the formation
of communities in collaboration networks is an intrinsically multiplex
process, which is the result of the interplay between intra-layer and
inter-layer triadic closure. For instance, in the case of scientific
collaborations, multiplex communities naturally arise from the fact
that scientists may collaborate with other researchers in their
principal field of investigation and with colleagues coming from other
scientific disciplines. Analogously, actors can prefer either to
specialise in a specific genre or instead to explore different
(sometimes dissonant) genres, and these two opposite behaviours
undoubtedly have an impact on the kind of meso-scale structures
observed on each of the layers of of the system. The generative model
we propose here mimics two of the most basic processes that drive the
evolution of collaborations in the real world, namely intra- and
inter-layer triadic closure, and is able to explain the appearance of
overlapping modular organisations in multi-layer systems. We will show
that the model is able to reproduce the salient micro-, meso- and
macro-scale structure of different real-world collaboration networks,
including the multi-layer network of co-authorship in journals of the
American Physical Society (APS) and the multiplex co-starring graph
obtained from the Internet Movie Database (IMDb).

\section*{Results}
\subsection*{Empirical analysis}
\label{sect:empirical}

We start by analysing the structure of two multiplex collaboration
networks from the real world. The first multiplex is constructed from
the APS co-authorship data set, and consists of four layers
representing four sub-fields of physics (respectively, Nuclear
physics, Particle physics, Condensed Matter I, and Interdisciplinary
physics). In particular, we considered only scientists with at least
one publication in each of the four sub-fields, and we connected two
scientists at a certain layer if they had co-authored at least a paper
in the corresponding sub-field. The second multiplex is constructed
from the Internet Movie Database (IMDb) and consist of four layers
respectively representing the co-starring networks of actors
  with at least one participation in four different genres, namely
  Action, Crime, Romance, and Thriller movies.  The basic structural
properties of each layer of the two multiplexes are summarised in
Table~\ref{tab:exAPSIMDbsingle} (see Methods for more information
about the data sets).

\begin{table}[ht]
\centering
\begin{tabular}{|l|lll|}
\hline
\textbf{APS} &  $N$ & $\langle k \rangle$ & $C$ \\
\hline
\hline
Nuclear (N) & 1238 & 4.75 & 0.27  \\
\hline
Particle (P) & 1238 & 4.66 & 0.30  \\
\hline
Cond. Matt. I (CM) & 1238 & 10.29 & 0.24  \\
\hline
Interdisciplinary (I) & 1238 & 7.37 & 0.26  \\
\hline
\hline
   \hline
   \textbf{IMDb} & $N$ & $\langle k \rangle$ & $C$  \\
\hline
\hline
  Action (A) & 55797 & 83.56 & 0.61 \\
   \hline
  Crime (C) & 55797 & 82.30 & 0.58 \\
   \hline
   Romance (R) & 55797 & 86.00 & 0.59 \\
   \hline
   Thriller (T) & 55797 & 77.75 & 0.56 \\
   \hline
\end{tabular}
\caption{\label{tab:exAPSIMDbsingle}\textbf{Basic properties of
    real-world multiplex collaboration networks.} We report the number
  of nodes $N$, the average degree $\langle k \rangle$ and the
  clustering coefficient $C$ for each layer of a subset of the APS
  and IMDb data sets. In particular, we focus on the multiplex
  collaboration network of all scientists active in Nuclear, Particle,
  Condensed Matter I and Interdisciplinary physics, and the multiplex
  collaboration network of all actors starring in Action, Crime,
  Romance and Thriller movies. All the layers of APS have a clustering
  coefficient $C$ in the range $[0.24,0.30]$. Conversely, the values of
  $C$ of all the IMDb layers are in the range $[0.56,0.61]$.}
\end{table}

Since we are interested in assessing the role of intra- and
inter-layer triadic closure in the formation of meso-scale multiplex
structures, we quantified the transitivity of each layer through the
clustering coefficient $C$~\cite{watts1998collective}, which takes values
  in the interval $[0,1]$ (see Methods). We notice that the four
layers of each data set have similar values of clustering, ranging
respectively in $[0.24,0.3]$ in the case of APS and in $[0.56,0.61]$
for IMDb. As we will discuss in the following, by focusing on layers
having comparable clustering we will be able to perform a
comparison between the structure of these real-world multiplex networks and the proposed model in its simplest formulation.

\begin{figure*}

\centering
\includegraphics[width=\linewidth]{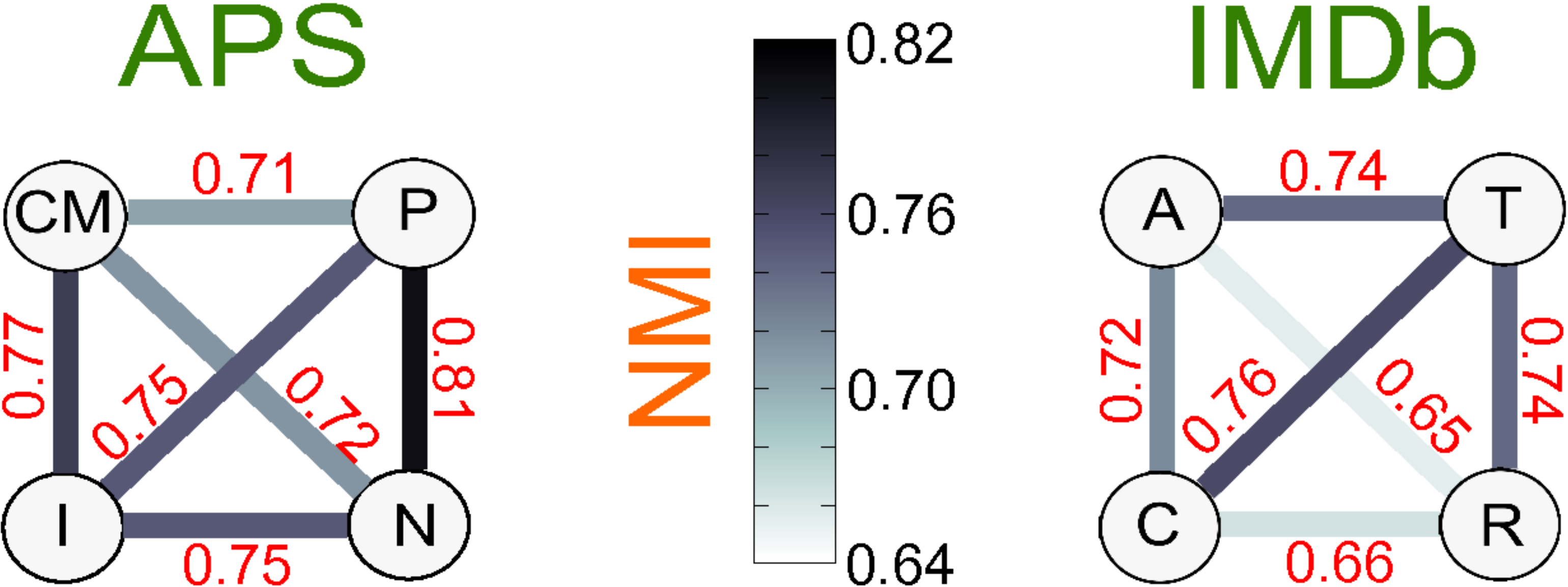}
\caption{\textbf{Similarity of communities at the different layers of
    real-world collaboration networks.}  In each of the two graphs
  nodes represent the layers of the multiplex (APS on the left and
  IMDb on the right) and the edges are coloured according to the value
  of the normalised mutual information for the community decompositions
  at the corresponding pairs of layers.}
\label{fig:fig1}
\end{figure*}

The multiplex nature of communities in collaboration networks can be
measured by means of the normalised mutual information (NMI)
\cite{danon2005comparing} (see Methods), which quantifies the similarity between
the partition in communities observed in two different layers of a
multiplex. The normalised mutual information takes values in
  $[0,1]$. In general, higher values of NMI correspond to more
similar partitions. The values of NMI for each pair of layers in APS
and IMDb are shown in Fig.~\ref{fig:fig1}. It is interesting to notice
that in general pairs of layers corresponding to related subjects or
genres exhibit higher values of NMI. This is for instance the case of
Nuclear Physics and Particle Physics in APS. Similarly, in the IMDb
network we observe a higher similarity between the communities at the
three layers representing respectively Thriller, Crime and Action
genres. Conversely, the layer of Romance movies displays a different
modular structure from Crime and Action. Notice also that the level of
similarity between the communities of two layers can vary
substantially, despite the four layers of each multiplex have roughly
the same clustering coefficient.

\subsection*{Model}
  In the following Section we introduce a model to grow
  collaboration networks with tunable multiplex community structure,
  able to reproduce the patterns observed in the considered real-world
  systems.  Let us consider for simplicity the case of a multiplex
  with $M=2$ layers, and assume that initially each layer consists of
  a clique of $n_0$ nodes. Then at each time step $t$ a new node is
  added to the network, with $m^{[1]}$ edge stubs to be connected on
  layer $1$ and $m^{[2]}$ other stubs to be connected on layer
  $2$. The multiplex network grows according to the following rules: 

\begin{itemize}
\item \emph{Layer selection.} The newly arrived node $i$
  selects one of the two layers $\{1,2\}$ uniformly at random. Let us label the
  first selected layer with the index $a$. The first
  edge of $i$ is connected to one of the existing nodes on that layer,
  chosen uniformly at random, that we call $n_a$.
\item \emph{Intra-layer triadic closure (I).} The remaining
  $m\lay{a}$-1 edges of node $i$ on layer $a$ are attached
  with probability $p\lay{a}$ to one of the first neighbours of
  $n_a$, chosen uniformly at random, and with probability
  $1-p\lay{a}$ to one of the nodes of layer $a$, chosen
  uniformly at random.
\item \emph{Inter-layer triadic closure.} When all its
  $m\lay{a}$ edges on layer $a$ have been created, node $i$
  starts connecting on the other layer
  $b$ with $m\lay{b}$ edges. The first link in layer $b$ is created with
  probability $p^*$ to the same node $n_a$, and with probability $1-p^*$ to one of
  the other nodes, chosen uniformly at random. The node to
  which this first link is attached is called $n_b$.
\item \emph{Intra-layer triadic closure (II).}  The remaining
  $m\lay{b}$-1 links at layer $b$ are attached with
  probability $p\lay{b}$ to one of the first neighbours of $n_b$
  chosen uniformly at random, and with probability $1-p\lay{b}$
  to one of the nodes at layer $b$, chosen uniformly at random.
\end{itemize}

\begin{figure*}
\centering 
\includegraphics[width=\linewidth]{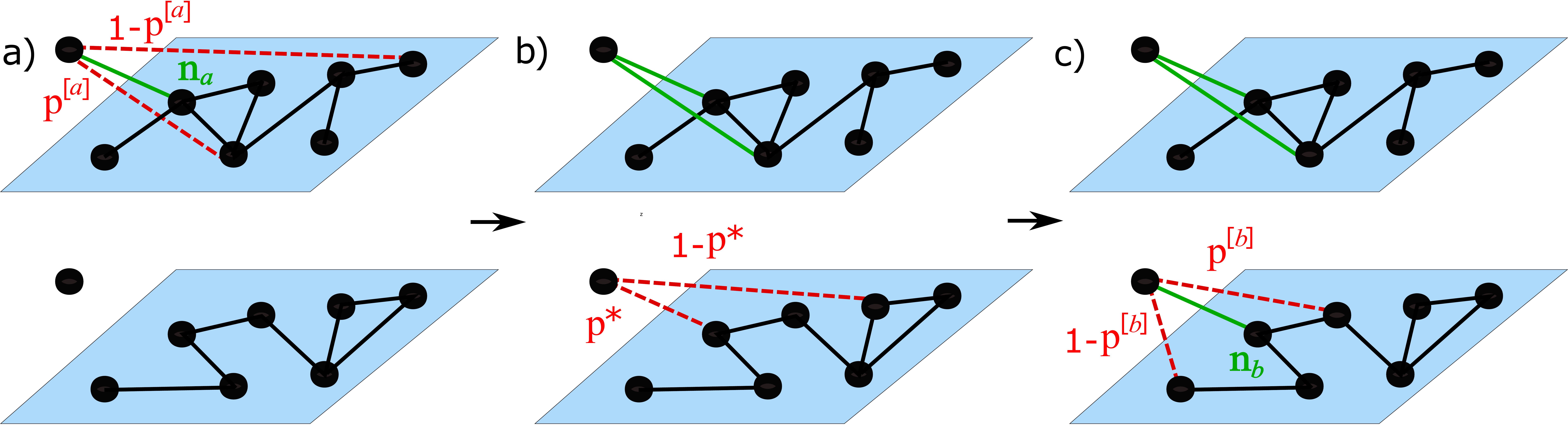}
\caption{\textbf{Schematic representation of network growth
      with intra-layer and inter-layer triadic closure.} A
    newly arrived node $i$ creates $m\lay{1}$ new edges on layer $1$
    and $m\lay{2}$ new edges on layer $2$. The new node starts by
    choosing at random one of the two layers $\{1,2\}$. We indicate the first chosen layer using the label $a$. a) The first link of the new node is connected to         one of the
    nodes of layer $a$, chosen uniformly at random and called $n_a$
    (solid green line). Each of the remaining $m\lay{a} - 1$ links is
    attached with probability $p\lay{a}$ to a neighbour of the
    previously chosen node (intra-layer triadic closure) or with
    probability $1-p\lay{a}$ to one of the nodes at layer $a$, chosen
    uniformly at random (dashed red lines). b) Afterwards, the new node starts connecting on the other layer $b$. The first link on layer
    $b$ is created to node $n_a$ with probability $p^*$, or to one of
    the other nodes at layer at random with probability $1-p^*$. We call $n_b$ the first
    node to which $i$ attaches on layer $b$. c) Each of the
    $m\lay{b}-1$ remaining edges on layer $b$ are attached with
    probability $p\lay{b}$ to one of the neighbours of $n_b$, and with
    probability $1-p\lay{b}$ to one of the nodes on layer $b$, chosen
    uniformly at random.}
\label{fig:fig2}
\end{figure*}

  This general model has five parameters to be tuned, namely the
  number of new edges $m\lay{1}$ and $m\lay{2}$ brought by each new
  node on each of the two layers, which determine the average degree
  on each layer, and the three probabilities $p\lay{1}$, $p\lay{2}$,
  and $p^*$, which are respectively responsible for the formation of
  intra- and inter-layer triangles. In fact, by varying the parameters
  $p\lay{1}$ and $p\lay{2}$ we can tune the strength of the intra-layer
  triadic closure mechanism, i.e the probability to form triangles on
  each of the two layers. In particular, larger values of
  $p\lay{1}$ and $p\lay{2}$ will foster the creation of a larger number of
  triangles in layer $1$ and layer $2$ respectively.  Conversely, the parameter $p^{*}$ tunes
  the inter-layer triadic closure mechanism, and in particular high
  values of $p^*$ correspond to a higher probability that the
  neighbourhoods of node $i$ at the two layers will exhibit a certain
  level of overlap.  These two simple attachment rules, namely intra-layer and inter-layer triadic 
  closure, aim to describe the real mechanisms characterising the evolution of collaboration networks.
  We argue that, for instance, scientists do not tend to
  collaborate with other scientists at random. Instead, they usually
  exploit the neighbourhoods of their collaborators in a specific
  field (\textit{intra-layer} triadic closure). Similarly, when
  opening themselves to new scientific fields, a researcher usually
  takes into account the neighbourhoods of their past colleagues from
  previous collaborations in other fields (\textit{inter-layer}
  triadic closure). A schematic representation of the model is
  depicted in Fig.~\ref{fig:fig2}.

  It has been recently shown~\cite{bianconi2014triadic} that in a
  single-layer network scenario the interplay between random
  attachment and triadic closure leads to a network growth in which
  the attachment probability (i.e., the probability for an existing
  node to receive one of the new edges) is a sub-linear function of
  the degree, and produces networks with non-trivial community
  structure, as long as the link density is not too high.  In the
  multi-layer model we propose, the further addition of an inter-layer
  triadic closure mechanism allows to tune at will the overlap between
  the community structures at the different layers.

\section*{Validation in a Simple Scenario}
\label{sec:3}
  To assess the ability of the model to reproduce the
  organisation of communities in multiplex networks, we start by
  considering a simple scenario, i.e. the case in which the layers of
  the multiplex have the same density ($m^{[1]}=m^{[2]}=m$) and the
  same clustering coefficient ($p^{[1]}=p^{[2]}=p$). We show that this
  simplified version of the model is already able to reproduce both
  the different levels of similarity between community structures at
  different layers, and the microscopic patterns of intra-layer and
  inter-layer degree correlations observed in the real-world
  collaboration multiplexes of APS and IMDb.

\begin{figure*}
\centering
\includegraphics[width=\linewidth]{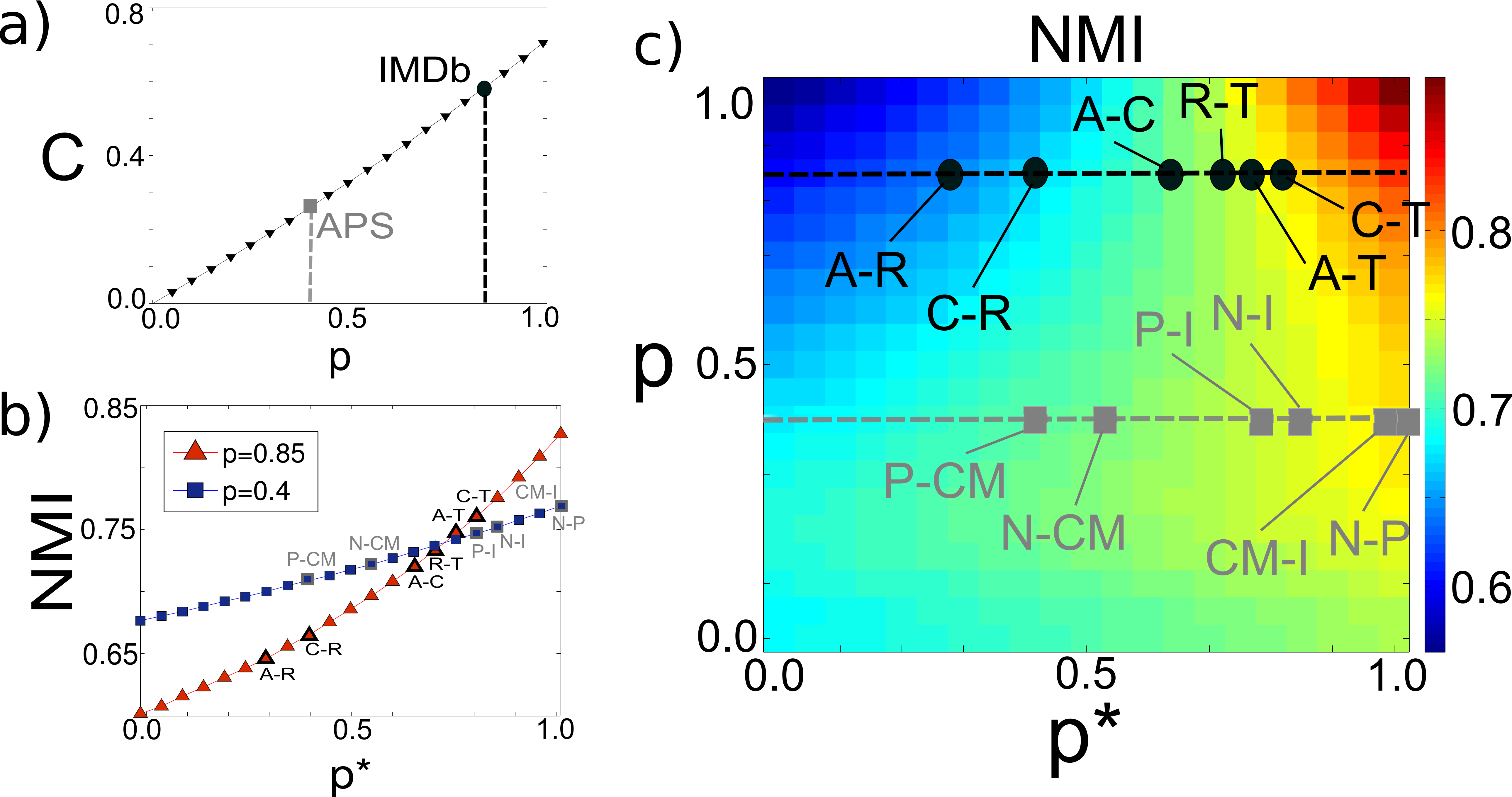}
\caption{\textbf{Model calibration in a simple scenario.} We
    show the values of $p$ and $p^*$ extracted for the different pairs
    of layers of the four-layer collaboration networks of APS and
    IMDb. (a) The clustering coefficient $C$ depends exclusively on
    the parameter $p$, which tunes intra-layer triadic closure. Since
    all the layers of those two multiplex networks have comparable
    clustering coefficients, we are able to determine the value of the
    parameter $p$ in each of the two cases. (b) For each pair of
    layers, we can also determine the value of the inter-layer triadic
    closure parameter $p^*$ by setting it equal to the value which
    yields an organisation in communities characterised by a value of
    NMI compatible with that observed in the real network.}
\label{fig:fig3}
\end{figure*}

  In Fig.~\ref{fig:fig3}(a), we report the values of the
  clustering coefficient $C$ (which, by construction, does not depend
  on the parameter $p^*$) for several realisations of the model (see
  Methods). As expected, the clustering coefficient of each layer is a
  linearly increasing function of the parameter $p$, which tunes the
  strength of intra-layer triadic closure. This means that, if we
  consider a real-world multiplex network whose layers have
  approximately the same value of clustering coefficient $C$, we can
  set the value of the parameter $p$ of the model
  accordingly. This is for instance the case of the
  four-layer multiplex networks of APS and IMDb constructed in the
  previous Section, where all the layers have comparable levels of
  clustering. We obtain $p=0.40$ for APS and $p=0.85$ for IMDb,
  respectively.

In Fig.~\ref{fig:fig3}(c) we show, as a colour-map, the values of NMI
of the networks obtained through the proposed model by using different
combinations of the parameters $p$ and $p^*$ (see Methods). It is
evident that, in spite of its simplicity, the model can yield a quite
rich variety of multiplex networks. In agreement with intuition, when
both $p$ and $p^*$ are large one obtains multiplexes with higher
values of NMI. In fact, in this regime both the intra-layer and
inter-layer triadic closure mechanisms are strongly affecting the
network evolution and, as a consequence, it is likely that the new
node joining the network will close a triad on both layers in the same
region of the network. As a consequence, each layer will have a strong
community structure (large $p$) which is pretty much correlated to the
one present on the other layer, due to the large value of inter-layer
triadic closure $p^*$. Conversely, if the inter-layer parameter $p^*$
is small we will obtain layers whose partitions in communities are
poorly correlated when $p$ is large (blue region in the phase space of
Fig.~\ref{fig:fig3}, while the NMI is only marginally larger when $p$
is very small (bottom-left corner of the phase space).

\begin{figure*}
\centering
\includegraphics[width=\linewidth]{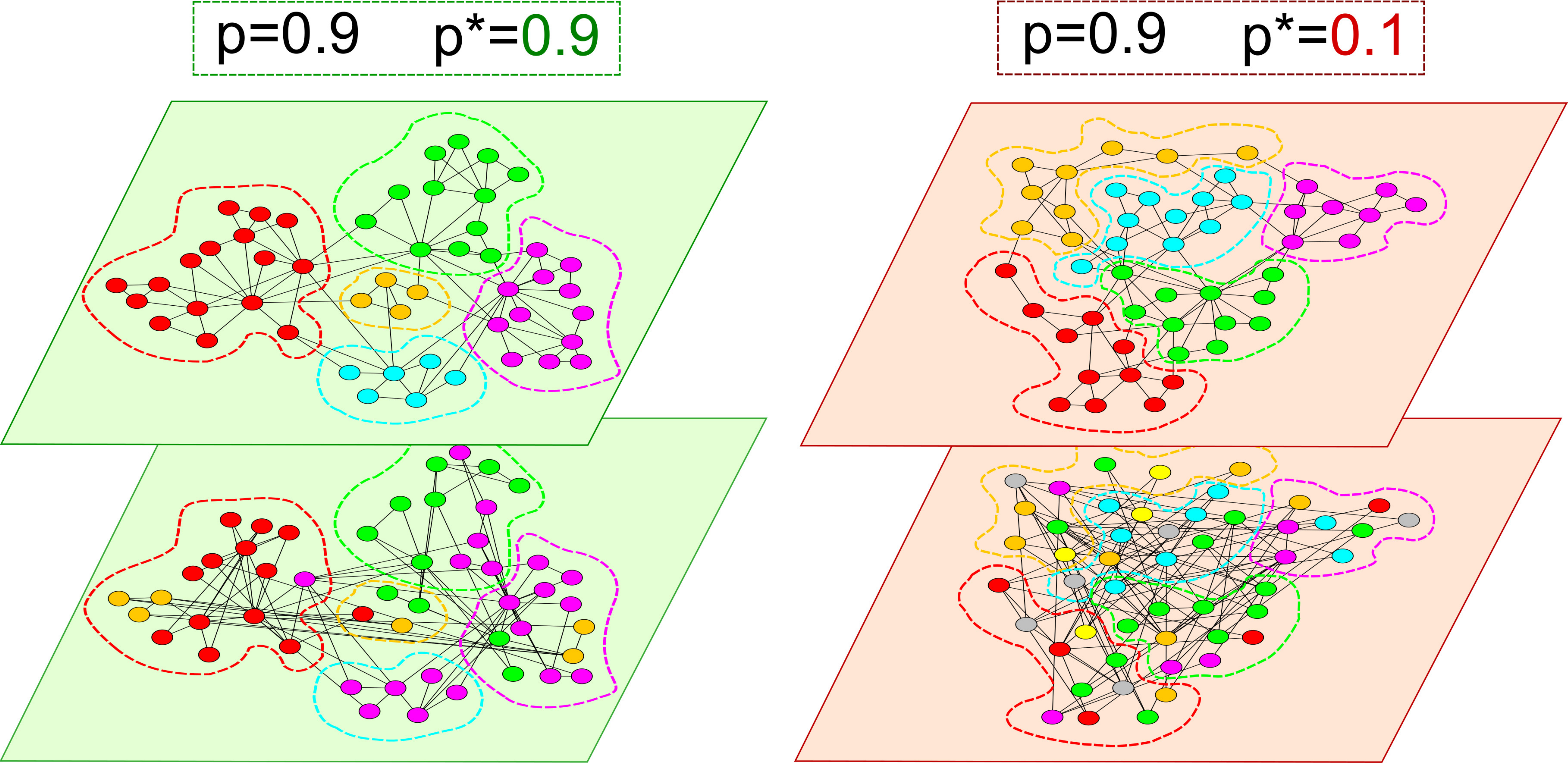}
\caption{\textbf{Layers with similar or dissimilar community
    structures.} We show the effect of the value of the inter-layer
  triadic closure parameter $p^*$ on the multiplex community
  structure. The two top layers show two typical realisations of the simplest version of the network model with $N=50$, $m^{[1]}=m^{[2]}=2$ and $p^{[1]}=p^{[2]}=0.9$. Nodes belonging to the same community are given the same colour and are drawn close to each
  other. The two layers at the bottom of each multiplex are obtained
  by setting, respectively, $p^*=0.9$ (left) and $p^*=0.1$
  (right). The nodes maintain the same placement in space on the
  second layer, but are coloured according to the community they
  belong in that layer (colours are chosen in order to maximise the
  number of nodes that have the same colour in the two layers). It is
  evident that the community structures of the two layers on the left,
  corresponding to $p^*=0.9$, are very similar, while the partition
  into communities of the upper layer on the left panel is
  substantially different from the one observed in the bottom layer of
  that multiplex.}
\label{fig:fig4}
\end{figure*}

  In Fig.~\ref{fig:fig4} we report two realisations of the
  multiplex network model with $N=50$, $m^{[1]}=m^{[2]}=2$ and
  $p^{[1]}=p^{[2]}=0.9$, respectively for $p^*=0.9$ (left) and
  $p^*=0.1$ (right). Nodes belonging to the same community are
  reported using the same colour, and the colour chosen for each
  community in the second layer (bottom) corresponds to the colour of
  the community in the first layer (top) for which the node overlap
  between the communities is maximum. These two examples help
  explain the role of the parameter $p^{*}$ in shaping the
  inter-layer modular structure of the network. For $p^*=0.9$ (left
  panel) the community structures of the two layers are closely
  matched (this situation corresponds to the high values of NMI found
  in the top-right region of the heat-map in Fig.~\ref{fig:fig3}),
  while for $p^*=0.1$ (right panel) the communities at the two layers
  are uncorrelated (low values of NMI in the top-left of the
  heat-map in Fig.~\ref{fig:fig3}).

Differently from the clustering coefficient $C$, the values of the
normalised mutual information NMI depend on both $p$ and $p^*$. Having
already determined a candidate value of $p$ for each multiplex by
fitting the clustering coefficient of its layers, we can determine the
strength of the inter-layer triadic closure mechanism by fitting the
NMI. Remarkably, for any fixed value of $p$, the simplest formulation
of our model is able to reproduce all the values of NMI observed in
the real-world networks by just tuning the parameter $p^*$, with the
exception of the pair Nuclear-Particle physics which is slightly out
of the plane with an NMI value of 0.81 (represented on the right
border of the plane which corresponds to NMI=0.79). We would
  like to note here that the model is able to produce a remarkably
  wide range of values of NMI, which span the whole interval $[0.6,
    0.9]$. 

   We further validate the model by showing that, using the
  inferred parameters ($p$,$p^*$), we are able to reproduce quite well
  the patterns of degree-degree correlations observed in the
  real-world collaboration multiplexes.

\begin{figure*}
\centering
\includegraphics[width=\linewidth]{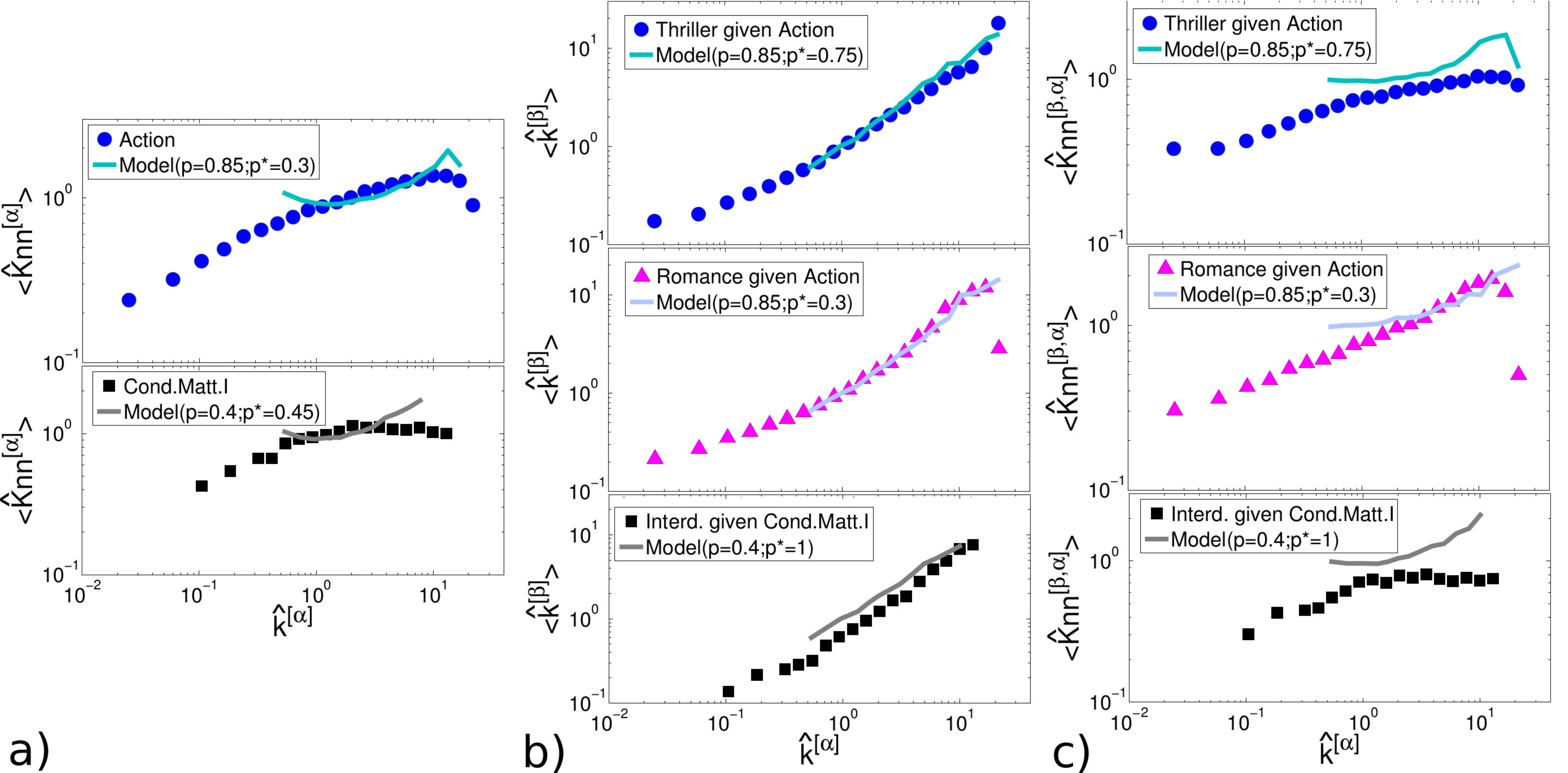}
\caption{\textbf{Intra-layer, inter-layer and mixed assortativity in
    collaboration networks.} We show the intra-layer (a), inter-layer
  (b) and mixed (c) degree-correlations for couples of layers of the
  IMDb and APS collaboration networks. Real data (dots) are compared
  with the results of our model (solid lines) generated with the
  extracted values $p$ and $p^*$. The symbols ($\>\hat{\>}\>$) indicate
  that the reported quantities (both for the model and the data) have
  been normalised to the values observed in the corresponding
  configuration model. As shown, the model is in general able to
  correctly capture the assortative trends of the three different
  types of correlations. Very good agreement with the data
  is attained in the case of the movie actor collaboration network. Less precise
  results are obtained for the APS network, where we deal with a
  system of considerably smaller size.}
\label{fig:fig5}
\end{figure*}

Indeed, for each pair of layers $\alpha$ and $\beta$ we analysed:
\begin{enumerate}
\item the intra-layer degree correlations, by looking at the average
  degree $\langle Knn^{[\alpha]}\rangle$ of the first neighbours on 
  layer $\alpha$ of nodes having a certain degree $k\lay{\alpha}$, as
  a function of $k\lay{\alpha}$;
\item the inter-layer degree correlations, by looking at the average
  node degree $\langle k^{[\beta]}\rangle$ on $\beta$ given the degree
  $k^{[\alpha]}$ on $\alpha$;
\item the mixed degree correlations by looking at the average mixed
  node degree $\langle Knn^{[\beta,\alpha]}\rangle$ given the degree
  $k^{[\alpha]}$ on $\alpha$;
\end{enumerate}    
(see Methods for details). The results are shown in
Fig.~\ref{fig:fig5} for some significant examples. Dots represent the
values measured on the real-world networks, while solid lines
correspond to the values obtained in the corresponding
multiplex models. Symbols with a hat ($\>\hat{\>}\>$) indicate that the
value of the considered variables, for both the model and the data,
have been normalised to the values of the corresponding configuration
model to allow a comparison (see Methods). It is interesting to notice
that the model reproduces quite well the three types of degree
correlations in the IMDb multiplex, both in the case of high $p$ and
high $p^*$ (Action, Thriller given Action) and the case of small
$p$ and small $p^*$ (Action, Romance given Action). A
  quantitative comparison of the the power-law fits of the curves is
  reported in Table~\ref{tab:exponents}. As an example from APS we
consider Condensed Matter I and Interdisciplinary physics (small $p$
and high $p^*$). In this case we observe marked differences in the
correlations measured in the real-world network and in the model
network, for both $\langle Knn^{[\alpha]}\rangle$ and $\langle
Knn^{[\beta,\alpha]}\rangle$. In particular, the model seems to
overestimate degree correlations. These discrepancies are probably due
to the relatively small number of nodes (only 1238) in the considered
data subset.

  Although our intention was not to exactly reproduce all the
  features observed in real-world collaboration multiplex networks, it
  is interesting to observe that the two mechanisms of inter-layer and
  intra-layer triadic closure play an important role in determining
  the degree-degree correlations in such networks. We also notice that
  the degree distributions of the layers in the synthetic networks are
  compatible with the stretched exponential functional forms
  introduced and discussed in Ref.~\cite{bianconi2014triadic}.

%

\begin{table}[]
\centering
\begin{tabular}{l|ll|ll}
Layers' pair & $\langle k^{[\beta]}\rangle$      &         & $\langle Knn^{[\beta,\alpha]}\rangle$    &         \\ \hline
           & $\gamma_{\rm data}$ & $\gamma_{\rm model}$ & $\gamma_{\rm data}$ & $\gamma_{\rm model}$ \\ \cline{2-5} 
Interd. given Cond. Matt. I       & 0.98      & 0.85       & 0.14      & 0.30       \\
Thriller given Action        & 0.83      & 0.84       & 0.16      & 0.17       \\
Romance given Action         & 0.89      & 0.87       & 0.30      & 0.27      
\end{tabular}
\caption{Quantitative comparison between the
    curves obtained from the model and the data for the inter-layer
    degree correlations the and mixed degree correlations. The
    curves have been fitted using a function of the form $f(x)\sim
    x^\gamma$; the $\gamma$ parameter is reported for the
    corresponding curves in Fig. \ref{fig:fig5}.}
\label{tab:exponents}
\end{table}

\section*{Model Calibration for Generic Multiplex Networks}
\label{sec:4}
  We now discuss how to calibrate the model in the most general
  case in which the layers might possibly have different edge density,
  i.e. $m^{[1]} \neq m^{[2]}$, and different clustering, i.e. $p^{[1]}
  \neq p^{[2]}$. As an example, we consider the co-authorship networks
  of the same four sub-fields of physics (namely, Nuclear, Particle,
  Condensed Matter I and Interdisciplinary physics) used to construct
  the four-layer APS multiplex (cf. Table~\ref{tab:exAPSIMDbsingle}
  and Fig~\ref{fig:fig1}). However, we focus here on all two-layer
  multiplex networks obtained by combining two networks at a time, so
  that, for instance, a node appears in the Nuclear-Particle (N-P)
  multiplex network if the corresponding author has published papers
  in both sub-fields. In general, the obtained multiplex networks are
  composed by layers with different edge density and different
  clustering coefficients, as shown in Table~\ref{tab:tabA}, thus we
  need to set separately the four parameters of the model $p\lay{1}$,
  $p\lay{2}$, $m\lay{1}$ and $m\lay{2}$.

\begin{figure*}
\centering
\includegraphics[width=\linewidth]{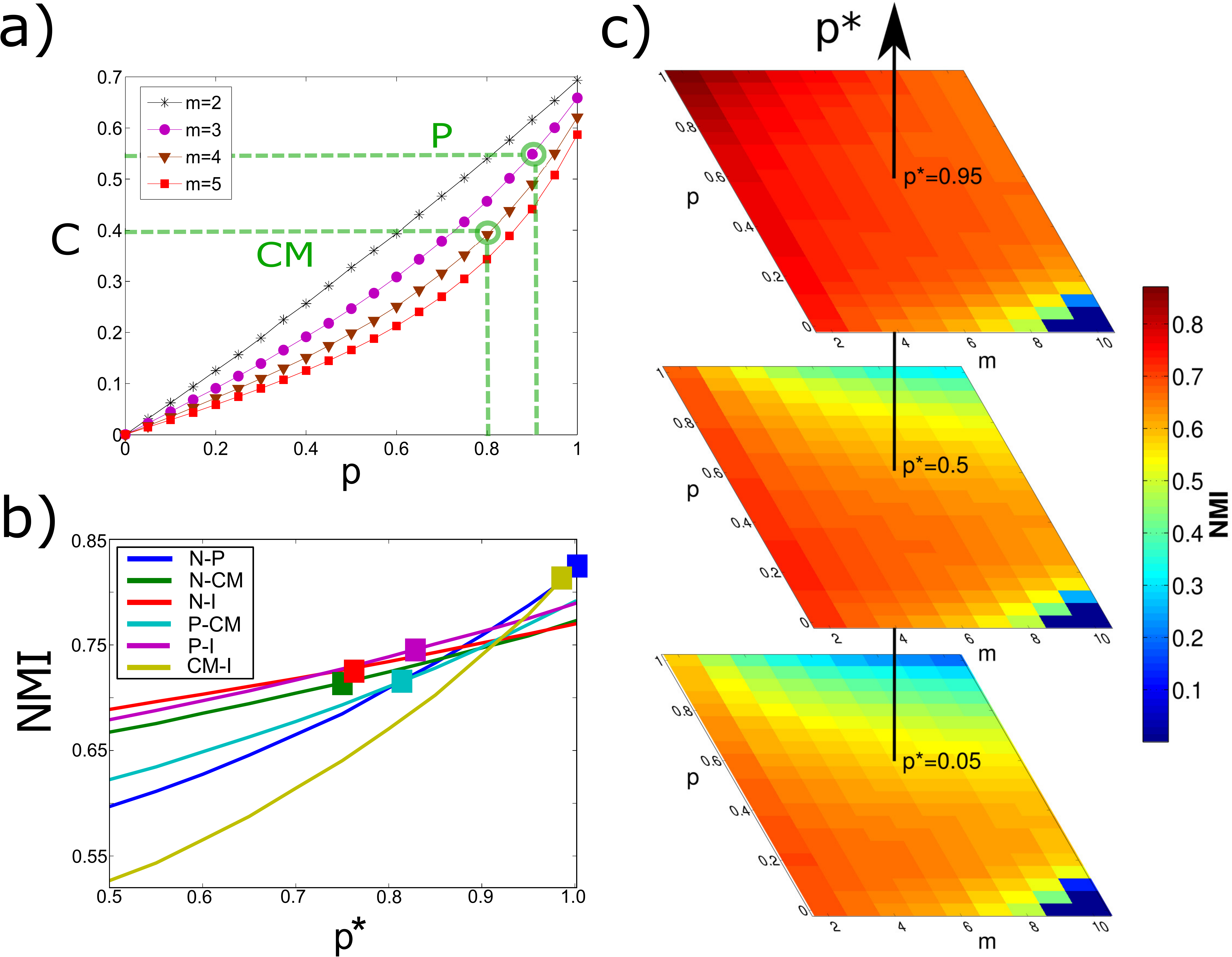}
\caption{\textbf{Model calibration.} In panel a) we
    show the dependence of the clustering coefficient $C$ on the
    intra-layer triadic closure parameter $p$ for different values of
    the parameter $m$, which sets the layer's average degree. In the
    multiplex consisting of the layers Particle (P) and Condensed
    Matter I (CM), the average degree of each layer corresponds,
    respectively, to $m\lay{1}=3$ and $m\lay{2}=4$. The value of
    $p\lay{1}$ and $p\lay{2}$ are determined to match the clustering
    coefficients $C\lay{1}$ and $C\lay{2}$. In panel b), after having
    determined $m^{[1]}$, $m^{[2]}$, $p^{[1]}$ and $p^{[2]}$ for all the
    pairs of layers in the APS dataset, we run the model with such
    parameters for different value of $p^*$ and infer, for each pair, the value of the inter-layer triadic closure parameter
    $p^*$ yielding a value of NMI compatible with that observed (see
    Table \ref{tab:exAPSIMDbsingle} for layers' acronyms).  In panel c)
    we plot a heat-map of the NMI as a function of $p$ and $m$,
    respectively for low (0.05), intermediate (0.50) and high (0.95)
    values of $p^*$ in the model with $m^{[1]}=m^{[2]}=m$ and
    $p^{[1]}=p^{[2]}=p$. An increase in the link density of the layers
    produces a less correlated community structure in the two layers,
    even if the inter- and intra-layer triadic closure strengths are
    high.}
\label{fig:figlast}
\end{figure*}

  We start by observing that the average degree of a synthetic
  layer is $ \avg{k} \simeq 2m$, where $m$ is the number of edge stubs
  connected by a newly arrived node, so that the parameters
  $m^{[1]}$, $m^{[2]}$ of the model can be set respectively equal to
  $\left[ \frac{\avg{k^{[1]}}}{2} \right]$ and $\left[
    \frac{\avg{k^{[2]}}}{2} \right]$, where $\avg{k\lay{1}}$ and
  $\avg{k\lay{2}}$ are the measured average degrees of the two layers (numbers are approximated to the closest integers). Similarly, as we show in Fig.~\ref{fig:figlast}(a), the
  clustering coefficient $C\lay{\alpha}$ of a layer $\alpha$ is
  univocally determined by $p^{[\alpha]}$, as soon as $m\lay{\alpha}$
  is fixed. In Fig.~\ref{fig:figlast}(a) we show how the values of
  $C\lay{\alpha}$ change as a function of $p\lay{\alpha}$, for
  different values of $m\lay{\alpha}$.  Hence, the values of the
  intra-layer triadic closure parameters $p\lay{1}$ and $p\lay{2}$ can
  be set in order to match the values of clustering coefficient
  observed in each of the two layers. The only parameter yet to be
  determined is $p^*$. However, if we set the values of $m\lay{1}$,
  $m\lay{2}$, $p\lay{1}$, and $p\lay{2}$ to match the densities and
  clustering coefficients of the layers, we can then run the model for
  different values of $p^*$ and look for the one which yields a value
  of NMI as close as possible to the one observed in the real
  two-layer multiplex. This procedure is sketched in
  Fig. \ref{fig:figlast} (b) for the six two-layer multiplexes in
  APS.

  In order to better understand the role of the different
  parameters, in Fig.~\ref{fig:figlast}(c) we report the values of NMI
  obtained from different realisations of the model with
  $m^{[1]}=m^{[2]}=m$ and $p^{[1]}=p^{[2]}=p$ for $m$ varying in
  $\left[2,3,...,10\right]$, and $p$ varying in
  $\left[0,0.1,...,1\right]$ at different values of $p^{*}$,
  $\left[0.05,0.5,0.95\right]$, corresponding respectively to low,
  intermediate and high inter-layer triadic closure strength. We see
  that the effect of the increase in the link density $m$ of the
  layers leads to a decrease in the similarity of their community
  structures even for high values of $p$ and $p^{*}$.

  It is interesting to notice that, although the generic version
  of the model depends on five parameters, respectively accounting for
  layer density ($m\lay{1}$ and $m\lay{2}$), triadic closure
  ($p\lay{1}$ and $p\lay{2}$), and inter-layer overlap of communities
  ($p^*$), the values of those parameters can be easily set by
  measuring just the average degree and the average clustering
  coefficient of each layer, and the normalised mutual information
  between the community structures at the two layers. Again, the good
  agreement between the synthetic networks and the real-world datasets extends also to other structural properties, such as
  intra-layer and inter-layer degree correlations, which were thought
  to have little or no direct relation at all with triadic
  closure. These results suggest that triadic closure plays an
  unexpectedly central role in determining the structural properties
  of real-world multiplex collaboration networks.

\begin{table}[ht]
\centering
\begin{tabular}{|l|l|l|l|l|l|l|l|}
\hline
\textbf{Layer 1} & \textbf{Layer 2} & $N$ &  $\langle k^{[1]} \rangle$ &  $\langle k^{[2]} \rangle$ &  $C^{[1]}$ &  $C^{[2]}$ & \textit{NMI} \\
\hline
\hline
N & P & 6572 & 6.88 &  7.46 &  0.56 &  0.56 & 0.83 \\
\hline
N & CM & 3828 & 4.53 & 7.20 &  0.43 &  0.34 & 0.71 \\
\hline
N & I & 2556 & 4.15 & 5.39&  0.37 &  0.33 & 0.72 \\
\hline
P & CM & 3774 & 5.70 & 7.82 &  0.53 &  0.40 & 0.71  \\
\hline
P & I & 2502 & 4.82 & 5.66 &  0.49 &  0.39 & 0.74  \\
\hline
CM & I & 27257 & 10.34 & 7.05 &  0.55 &  0.64 & 0.82 \\
\hline
\end{tabular}
\caption{\textbf{\label{tab:tabA} Basic properties of duplex networks
    in APS.} We consider all the possible multiplex networks with
  $M=2$ layers obtained from combinations of the APS collaboration
  networks corresponding to the four sub-fields Nuclear, Particle,
  Condensed Matter I and Interdisciplinary Physics. For each duplex, we
  report the number of nodes $N$, the average degree on the two layers
  $\langle k^{[1]} \rangle$ and $\langle k^{[2]} \rangle$, and the
  values of the clustering coefficients $C^{[1]}$ and $C^{[2]}$.}
\end{table}


\section*{Discussion}

Human collaboration patterns are inherently multifaceted and often
consist of different interaction layers. Scientific collaboration is
probably the most emblematic example. As a Ph.D. student you usually join the scientific collaboration network by publishing the first paper
with your supervisor in a specific field. Afterwards, you start being
introduced by your supervisor to other researchers in the same field,
e.g. to some of his/her past collaborators, and you might end up working
with them, creating new triangles in the collaboration network of your
field (what we called intra-layer triadic closure). But it is also
quite probable that some of your past collaborators will in turn
introduce you to researchers working in another -possibly related- area
(what we called an inter-layer triadic closure), so that you will
easily find yourself participating in more than just one field, and
the collaboration network around you will become multi-dimensional. Such multi-level collaboration patterns appear not to be
specific of scientific production only, but are instead found in many
aspects of human activity.

The multi-layer network framework provides a natural way of modelling
and characterising multidimensional collaboration patterns in a
comprehensive manner. In particular, we have argued that one of the
classical mechanisms responsible for the creation of triangles of
acquaintances, i.e. triadic closure, is indeed general enough to give
also account for another interesting aspect of multi-level
collaboration networks, namely the formation of cohesive communities
spanning more than a single layer of interaction. It is quite
intriguing that the simple model we proposed in this work, based just
on the interplay between intra- and inter-layer triadic closure, is
actually able to explain much of the complexity observed in the micro-
meso- and macroscopic structure of multidimensional collaboration
networks of different fields (science and movies), including not just
transitivity but also intra- and inter-layer
degree correlation patterns and the correspondence between the
community partitions at difference layers. We also remark that such
levels of accuracy in reproducing the features of real-world systems have been obtained
without the introduction of ad-hoc ingredients.

The results reported in this paper suggest that, despite the apparent
differences in the overall dynamics driving scientific cooperation and
movie co-starring, triadic closure is a quite generic mechanism and
might indeed be one of the fundamental processes shaping the structure
of multi-layer collaboration systems. These findings fill a gap in the
literature about modelling growing multidimensional networks, and pave
the way to the exploration of other simple models which can help
underpinning the driving mechanisms responsible for the emergence of
complex multi-dimensional structures.


\section*{Methods}

\textit{Data sets. --} We considered data from the APS and the IMDb
collaboration networks. The APS collaboration data set is available
from the APS website \texttt{http://journals.aps.org/datasets} in the
form of XML files containing detailed information about all the papers
published by all APS journals. The download is free of charge and
restricted to research purposes, and APS does not grant to the
recipients the permission to redistribute the data to third parties.
We parsed the original XML files to retrieve, for each paper, the list
of authors and the list of PACS codes. The PACS scheme provides a
taxonomy of subjects in physics, and is widely used by several
journals to identify the sub-field, category and subject of papers.
We used the highest level of PACS codes to identify the ten main
sub-fields of physics, and we considered only the papers published in
Nuclear physics, Particle physics, Condensed Matter I and
Interdisciplinary physics, respectively associated to high-level PACS
codes starting with 1 (Particle physics), 2 (Nuclear physics), 6
(Condensed Matter I) and 8 (Interdisciplinary physics).  We focused only
on the authors who had at least one publication in each of the four
sub-fields~\cite{nicosia2015measuring}. The co-authorship network of each of
those four sub-fields constitutes one of the four layers of the APS
multiplex. In particular, two authors are connected on a certain layer
only if they have co-authored at least one paper in the corresponding
sub-field. In the construction of the collaboration network of each
sub-field we purposely left out papers with more than ten authors,
which represent big collaborations whose driving dynamics might be
more complex than just triadic closure.

The IMDb data set is made available at the website
\texttt{ftp://ftp.fu-berlin.de/pub/misc/movies/database/} for personal
use and research purposes. The data set comes in the form of several
compressed text files, and we used those containing information about
actors, actresses, movies and genres. We focused only on the
co-starring networks of four movie genres, namely Action, Crime,
Romance, and Thriller~\cite{nicosia2015measuring}, obtained by merging
information about participation of actors and actresses to each
movie. In particular, two actors are connected by a link on a given
layer (genre) only if they have co-starred in at least one movie of
that genre.
We considered only the actors who had acted in at least one movie of
each of the four genres. We chose to restrict our analysis to just four
layers for both the APS and the IMDb data set, which allowed us to
consider the simplest formulation of our model, in which all the
layers have the same clustering coefficient $C$. The use of the APS
and the IMDb data sets does not require any ethical approval.

\textit{Transitivity and community structure. ---} We measured the
transitivity of each level by mean of the clustering coefficient
$C=(1/N)\sum_i C_i$ \cite{watts1998collective}, where $C_i$:
\begin{equation}
C_i = \frac{\sum_{j \neq i, m \neq i} a_{ij}a_{jm}a_{mi}}{\sum_{j \neq i, m \neq i} a_{ij}a_{mi}}=\frac{\sum_{j \neq i, m \neq i} a_{ij}a_{jm}a_{mi}}{k_i(k_i-1)}.
\end{equation}

The similarity of two community partitions can be measured through the
normalised mutual information (NMI)~\cite{danon2005comparing}.  In particular,
given the two partitions ${\cal P}_{\cal \alpha}$ and ${\cal P}_{\cal
  \beta}$ respectively associated to layer $\alpha$ and layer $\beta$,
we denote the normalised mutual information (NMI) between them as 
\begin{equation}
NMI({\cal P}_{\cal \alpha}, {\cal P}_{\cal \beta})  = \frac{-2\sum^{M_\alpha}_{m=1}\sum^{M_\beta}_{m'=1}
N_{m m'}\log\left(\frac{N_{m m'}N}{N_{m}N_{m'}}\right)}
{\sum^{M_\alpha}_{m=1}N_{m}\log\left(\frac{N_{m}}{N}\right)
 + \sum^{M_\beta}_{m'=1}N_{ m'} \log\left(\frac{N_{ m'}}{N}\right)}
\end{equation}
where $N_{m m'}$ is the number of nodes in common between module $m$
of partition ${\cal P}_{\cal \alpha}$ and module $m'$ of partition
${\cal P}_{\cal \beta}$, while $N_{m}$ and $N_{ m'}$ are respectively
the number nodes in module $m$ and in module $m'$.  The partition in
communities on each layer has been obtained through the algorithm
Infomap~\cite{rosvall2008maps}.

\textit{Synthetic multiplex networks. ---} We created synthetic
networks according to our multi-layer network model by starting, on
each layer, from a seed graph consisting of a triangle of nodes and
simulating the intra- and inter-layer triadic closure mechanism for
$N=20000$ nodes, for different values of the parameters $p$ and
$p^*$. For each pair of values $(p, p^*)$ we computed the mean
clustering coefficient $C$ on each single layer and the normalised
mutual information NMI of the community partitions of the two layers
over 30 different realisations. As observed from simulations, once the
parameters $(p,p^*)$ are fixed, the values of NMI and $C$ do not
vary substantially as the order $N$ of the network increases.  Notice
that since the most simple formulation of the model we have set an
identical value of $p$ on both layers, the two layers will end up
having the same clustering coefficient (up to small finite-size
fluctuation).

\textit{Degree correlations. ---} We study the assortativity of real
multiplex collaboration networks in terms of intra-layer, inter-layer
and mixed degree correlations. The trend for intra-layer correlations
is analysed by mean of the function $\langle
K^{[\alpha]}_{nn}(k^{[\alpha]}) \rangle$, that is the average degree
of the nearest neighbours on layer $\alpha$ of a node with given
degree $k\lay{\alpha}$ on that layer. In particular, $\langle
K^{[\alpha]}_{nn} \rangle$ is obtained as an average of
$K^{[\alpha]}_{nn, i}$ over all nodes with the same degree
$k^{[\alpha]}$. The node term can be computed as $K^{[\alpha]}_{nn,
  i}=\frac{\sum_{j \neq i}
  a^{[\alpha]}_{ij}k_j^{[\alpha]}}{k_i^{[\alpha]}}$, where
$a^{[\alpha]}_{ij}$ are the entries of the adjacency matrix at layer
$\alpha$. Since such measure considers only a layer at a time, the
layer index here is not strictly necessary but will be kept for
symmetry with the other coefficients.  It is interesting to notice
that, in absence of intra-layer degree correlations, $\langle
K^{[\alpha]}_{nn}(k^{[\alpha]}) \rangle$ is a constant, while $\langle
K^{[\alpha]}_{nn}(k^{[\alpha]}) \rangle$ is an increasing (resp.,
decreasing) function of $k\lay{\alpha}$ if assortative (resp.,
disassortative) degree correlations are present.

To quantify inter-layer degree correlations we considered the quantity
$\langle k^{[\beta]}(k^{[\alpha]})
\rangle$~\cite{nicosia2013growing,nicosia2015measuring}, that is the average degree on layer
$\beta$ of a node with degree $k^{[\alpha]}$ on layer $\alpha$. Again,
$\langle k^{[\beta]}(k^{[\alpha]}) \rangle$ will be an increasing
function of $k\lay{\alpha}$ if nodes tend to have similar degrees on
both layers (assortative inter-layer correlations), while $\langle
k^{[\beta]}(k^{[\alpha]}) \rangle$ will decrease with $k\lay{\alpha}$
if a hub on one layer will preferentially have small degree on the
other layer, and vice-versa.

Finally, we measured the presence of mixed correlations through the
function $\langle K^{[\beta, \alpha]}_{nn}(k^{[\alpha]}) \rangle$,
that is the average degree on layer $\beta$ of the nearest neighbours
on layer $\beta$ of a node with degree $k^{[\alpha]}$ on layer
$\alpha$~\cite{nicosia2014nonlinear}. In analogy with the case of intra-layer correlations, the node
term is $K^{[\beta,\alpha]}_{nn, i}=\frac{\sum_{j \neq i}
  a^{[\beta]}_{ij}k_j^{[\beta]}}{k_i^{[\alpha]}}$. We remark here that
there exists another possible definition of mixed correlations
coefficient, which considers the nearest neighbours of a node on layer
$\alpha$ rather then $\beta$ (see Ref.~\cite{nicosia2014nonlinear} for
details). The results for the alternative definition of mixed
correlations are analogous to those observed for $\langle K^{[\beta,
    \alpha]}_{nn}(k^{[\alpha]}) \rangle$ and are not shown in the
text.
In general, correlation functions might be affected by the degree sequence at each layer of the multiplex. In the simple scenario considered at first,
however, we do not fit the parameter $m$ from the data, to reduce as much as possible the complexity of the model.
Instead, in order to still perform an accurate comparison between the synthetic
multiplex networks constructed by our model and the real ones, in a second step we
divided all the correlation functions by their (constant) value
expected in the corresponding configuration model network. The correct
normalisation for the intra-layer correlation function is
$\frac{\avg{(k\lay{\alpha})^2}}{\avg{k\lay{\alpha}}}$~\cite{catanzaro2005generation},
while for the inter-layer correlation function we have to divide
$\langle k^{[\beta]}(k^{[\alpha]}) \rangle$ by
$\avg{k\lay{\beta}}$. Finally, the mixed correlation function is
correctly normalised by
$\frac{\avg{(k\lay{\beta})^2}}{\avg{k\lay{\alpha}}}$.

\section*{Acknowledgements}
F.B., V.N. and V.L. acknowledge support by the Project LASAGNE,
Contract No.318132 (STREP), funded by the European Commission.


\end {document}